\documentclass[aps,prd,onecolumn,groupedaddress,nofootinbib,amssymb]{revtex4}
\usepackage[dvips]{graphicx}
\usepackage{amssymb}
\usepackage{amsmath}
\usepackage{graphicx,color}
\usepackage{amsfonts}
\usepackage{bm}
\usepackage{cancel}
\usepackage{comment}

\newcommand\e{\mathrm{e}}

\usepackage{babel}

\allowdisplaybreaks[4]

\begin{document}

\tolerance=5000

\title{Generalized Black Hole Entropy in Two Dimensions}
\author{S.~Nojiri$^{1,2}$}
\email{nojiri@gravity.phys.nagoya-u.ac.jp}
\author{S.~D.~Odintsov$^{3,4}$}
\email{odintsov@ice.cat}
\author{V.~Faraoni$^5$}
\email{vfaraoni@ubishops.ca}

\affiliation{$^{1)}$ Department of Physics, Nagoya University, Nagoya 
464-8602, Japan \\
$^{2)}$ Kobayashi-Maskawa Institute for the Origin of Particles
and the Universe, Nagoya University, Nagoya 464-8602, Japan \\
$^{3)}$ ICREA, Passeig Luis Companys, 23, 08010 Barcelona, Spain \\
$^{4)}$ Institute of Space Sciences (ICE, CSIC) \\
C. Can Magrans
s/n, 08193 Barcelona, Spain \\
$^5$ Department of Physics \& Astronomy, 
Bishop's University, 2600 College Street, Sherbrooke, Qu\'ebec, Canada J1M 
1Z7
}

\begin{abstract}
The Bekenstein-Hawking entropy of a black hole is proportional to its 
horizon area, hence in $D=2$ spacetime dimensions it is constant because 
the horizon degenerates into two points. 
This fact is consistent with Einstein's gravity becoming topological in two dimensions. 
In $F(R)$ gravity, which is non-trivial even in $D=2$, we find that the entropy is constant, as for Bekenstein-Hawking. 
As shown in EPL 139 (2022) no.6, 69001 (arXiv:2208.10146), two-dimensional $F(R)$ gravity is 
equivalent to Jackiw-Teitelboim gravity, in turn  equivalent to the Sachdev-Ye-Kitaev model 
where the entropy becomes constant in the large $N$ limit. 
Several recently proposed entropies are functions of the Bekenstein-Hawking entropy and become constant in $D=2$, but in 
two-dimensional dilaton gravity entropies are not always constant. 
We study general dilaton gravity and obtain arbitrary static black hole 
solutions for which the non-constant entropies depend on the mass, horizon 
radius, or Hawking temperature, and constitute new proposals for a generalized entropy.

\end{abstract}

\maketitle

\section{Introduction}
\label{Sec1}

The Bekenstein-Hawking entropy~\cite{Bekenstein:1973ur, Hawking:1975vcx} of a black hole is 
proportional to the area of its event horizon. In two spacetime dimensions, this horizon 
degenerates into two points whose area vanishes and, therefore, the Bekenstein-Hawking 
entropy reduces to a constant independent of the black hole mass $M$, the horizon radius 
$r_\mathrm{H}$, or the Hawking temperature $T$. Of course, in $D=2$ spacetime dimensions, 
Einstein's gravity becomes trivial because the Ricci scalar $R$ reduces to a total 
derivative, and the Einstein-Hilbert action becomes a topological number. However, it has 
been shown that, by taking the limit of spacetime dimension two by adjusting the 
gravitational coupling, the black hole solution survives \cite{Nojiri:2020tph}. In this 
sense, as long as we consider Einstein's gravity, there is no conflict with the fact that the 
Bekenstein-Hawking entropy becomes constant. Here we first consider $F(R)$ 
gravity~\cite{Capozziello:2002rd, Nojiri:2003ft}, whose action is given by a function of $R$ 
and, by evaluating its action, we show that entropy is constant as for the Bekenstein-Hawking 
entropy. Motivated by early-universe inflation and by the late-time accelerating expansion of 
the universe, many modified gravity theories have been studied (see \cite{Capozziello:2009nq, 
Faraoni:2010pgm, Capozziello:2011et, Nojiri:2010wj, Nojiri:2017ncd} for reviews) and $F(R)$ 
gravity is probably the most popular class of theories for cosmological purposes. Recently it 
has been shown \cite{Nojiri:2022mfi} that, in two spacetime dimensions, $F(R)$ gravity is 
equivalent to Jackiw-Teitelboim (JT) gravity~\cite{Teitelboim:1983ux, Jackiw:1984je}. The 
latter has been studied intensively because it is equivalent to the Sachdev-Ye-Kitaev (SYK) 
model~\cite{Sachdev:1992fk, Kitaev}. The explicit relation between  JT gravity and the SYK 
model 
is discussed in \cite{Jensen:2016pah, Maldacena:2016upp}. It would be interesting 
if the SYK model could saturate the chaos bound \cite{Maldacena:2015waa, Shenker:2013pqa}. It 
has been shown that, even in the SYK model, entropy approaches a constant in the large $N$ 
limit \cite{Sarosi:2017ykf}. This finding is consistent with the result of the present work 
about black hole entropy in $F(R)$ gravity which, in two dimensions, is equivalent to  JT 
gravity. 
Horizon radius or temperature-dependent corrections to the Bekenstein-Hawking 
entropy would then seem natural.
 
Several new entropy proposals have been advanced recently, including the Tsallis 
entropy~(\cite{tsallis}, see also \cite{Ren:2020djc, Nojiri:2019skr}) for systems with 
long-range interactions, the R{\'e}nyi information entropy~\cite{renyi,Czinner:2015eyk, 
Tannukij:2020njz}, the Kaniadakis entropy generalizing the Boltzmann-Gibbs entropy in 
relativistic statistical systems \cite{Kaniadakis:2005zk}, the Sharma-Mittal 
entropy~\cite{SayahianJahromi:2018irq} related to information theory, the Barrow 
entropy~\cite{Barrow:2020tzx} proposed as a toy model for quantum spacetime foam, and the 
non-extensive Loop Quantum Gravity entropy of \cite{Majhi:2017zao,Czinner:2015eyk}. 
Furthermore, a new and very general entropy construction with the above properties and 
containing all these new entropy proposals as special limits was introduced in 
\cite{Nojiri:2022aof}. This generalized entropy contains six parameters, while a simpler 
alternative with three parameters was also proposed \cite{Nojiri:2022aof}. 
These generalized entropies, however, are functions of the Bekenstein-Hawking entropy, and they all 
become constant, and physically irrelevant, in two dimensions.

In the case of $D=2$ dilaton gravity (see \cite{Nojiri:2000ja} for a review), black hole 
entropies are well-known. For example, in \cite{Myers:1994sg} the black hole entropy in the 
Callan-Giddings-Harvey-Strominger (CGHS) model~\cite{Callan:1992rs} was evaluated using 
Wald's method \cite{Wald:1993nt} based on a Noether current. The entropy thus obtained 
differs from the Bekenstein-Hawking entropy and depends on the value of the dilaton. 
Therefore, entropy can depend on the black hole mass, the horizon radius (note that the 
Hawking temperature in the CGHS model is constant), and does not coincide with any of the 
above-mentioned generalized entropies ({\em e.g.}, the Tsallis entropy or the R{\'e}nyi 
entropy, which are functions of the Bekenstein-Hawking entropy). In the following, we 
investigate general dilaton gravity (not only the CGHS model) and obtain solutions describing 
arbitrary static spacetimes. For the models examined, the entropies $\mathcal{S}$ are 
general functions of the black hole mass $M$, the horizon radius $r_\mathrm{H}$, or the 
Hawking temperature $T$,
\begin{align} 
\label{entropy} 
\mathcal{S}=\mathcal{S}\left(M\right)\quad 
\mbox{or} \quad \mathcal{S} = \mathcal{S}\left(r_\mathrm{H}\right) \quad 
\mbox{or} \quad \mathcal{S}=\mathcal{S}(T)\,.
\end{align} 

The reconstruction method we follow is similar to that of Ref.~\cite{Nojiri:2022sfd}. The 
entropy~(\ref{entropy}) could constitute a new type of generalized entropy. In spacetime 
dimension $D>2$ the Bekenstein-Hawking entropy depends on the black hole mass $M$, the 
horizon radius $r_\mathrm{H}$, or the Hawking temperature $T$ and generalized (Tsallis, 
R{\'e}nyi, {\em etc.}) entropies can be regarded as general functions of $M$, $r_\mathrm{H}$, 
$T$. Therefore, the generalized entropies in two dimensions, if they exist, could also be 
general functions of these quantities. We show that such entropy always 
exists. 

In the next section, we study the entropy of two-dimensional $F(R)$ 
gravity and of JT 
gravity, reviewing these theories previously discussed in \cite{Nojiri:2022mfi}. In 
Sec.~\ref{subsec2-2}, we find a black hole solution of $D=2$~ $F(R)$ gravity even though no 
black hole solution exists in two-dimensional Einstein gravity. In Sec.~\ref{subsec2-3}, 
we calculate the entropy of this $F(R)$ black hole and show that it is constant, consistent 
with the SYK model. In Sec~\ref{Sec3}, we consider black hole entropies in general dilaton 
gravity with a kinetic term in two dimensions. Section~\ref{subsec3-1} explains, for an 
arbitrary static spacetime, how to reconstruct the dilaton gravity model which generates 
this solution: an explicit example is provided in Sec.~\ref{subsec3-2}. In 
Sec.~\ref{subsec3-3}, we 
calculate the entropy for this example and show that it is not constant but it depends on the 
black hole mass $M$, the horizon radius $r_\mathrm{H}$, or the Hawking temperature $T$. We 
then present a framework to obtain the entropy in general dilaton gravity as a function of 
$M$, $r_\mathrm{H}$, $T$. Sec.~\ref{Sec4} contains a summary and a discussion.

\section{$D=2$ $F(R)$ gravity and Jackiw-Teitelboim gravity} 
\label{Sec2}

Here we compute the thermal entropy of a black hole in $F(R)$ gravity in 
$D=2$ spacetime dimensions and in JT gravity.

\subsection{Equivalence between two-dimensional $F(R)$ gravity and 
Jackiw-Teitelboim gravity}
\label{subsec2-1}

It was shown in \cite{Nojiri:2022mfi} that, when the 
spacetime dimension is two, $F(R)$ gravity with action in general $D$ 
spacetime dimensions\footnote{The metric signature is 
${-}{+}{+}{+}$, $\kappa^2 \equiv 8\pi G$ where $G$ is Newton's constant, 
and we use units in which the speed of light $c$ and the reduced Planck constant $\hbar$ are 
unity.}
\begin{align}
\label{FR1}
S=\frac{1}{2\kappa^2} \int dx^D \sqrt{-g} \, F(R)\, ,
\end{align}
is equivalent to JT gravity with action 
\begin{align}
\label{WFR7B}
S=\frac{1}{2\kappa^2} \int dx^2 \sqrt{-g} \, \phi \left( R - \Lambda \right)\, . 
\end{align}
Let us briefly review the proof of this equivalence. The action 
(\ref{FR1}) of $D$-dimensional $F(R)$ gravity can be 
rewritten as
\begin{align}
\label{FR2}
S=\frac{1}{2\kappa^2} \int dx^D \sqrt{-g} \left[ \phi R - V(\phi) \right]\, ,
\end{align}
its variation with respect to $\phi$ gives 
$R=V'(\phi)$, which can be solved for the effective scalar field degree of freedom $\phi(R)$.
By substituting the result into the action~(\ref{FR2}), one obtains the 
action~(\ref{FR1}) with $F(R) = \phi(R) R - V\left(\phi\left(R\right) \right)$. 
The scale transformation $g_{\mu\nu}\to \e^\sigma g_{\mu\nu}$ then transforms the action~(\ref{FR2}) into
\begin{align}
\label{WFR2}
S=\frac{1}{2\kappa^2} \int dx^D \sqrt{-g} \, \e^{\frac{D}{2} \, \sigma} 
\left\{ \phi \left[ R - (D-2)\Box \sigma - \frac{(D-3)(D-2)}{4} \, \partial^\mu 
\sigma \partial_\mu \sigma \right] \e^{-\sigma} - V(\phi) \right\}\, .
\end{align}
Except for $D\neq 2$, the choice $\sigma$ such that 
$\e^{\frac{D-2}{2} \, \sigma}\phi=1$ reproduces the well-known scalar-tensor gravity
\begin{align}
\label{WFR4}
S=\frac{1}{2\kappa^2} \int dx^D \sqrt{-g} \left\{ R - \frac{(D-3)(D-2)}{4} \, \partial^\mu \sigma \partial_\mu \sigma
 - \e^{\frac{D}{2}\sigma} \, V\left( \phi=\e^{- \frac{D-2}{2} \, \sigma} \right) \right\}\,,
\end{align}
therefore scalar-tensor theory is equivalent to $F(R)$ gravity 
\cite{Maeda:1988ab}. If, instead, $D=2$ the action~(\ref{WFR2}) has the form 
\begin{align}
\label{WFR5}
S=\frac{1}{2\kappa^2} \int dx^2 \sqrt{-g} \left[ \phi R - \e^\sigma V(\phi) \right] \, .
\end{align}
By choosing $\sigma$ so that $\e^\sigma V(\phi) = \phi \Lambda$ with  $\Lambda$ constant, the 
action~(\ref{WFR5}) becomes
\begin{align}
\label{WFR7B}
S=\frac{1}{2\kappa^2} \int dx^2 \sqrt{-g} \, \phi \left( R - \Lambda 
\right)\, ,
\end{align}
which is nothing but the action of JT gravity~\cite{Teitelboim:1983ux, Jackiw:1984je}. 
Therefore, at the classical level,  vacuum $F(R)$ gravity in two dimensions is equivalent 
to JT gravity.

\subsection{Black hole in two-dimensional $F(R)$ gravity}
\label{subsec2-2}

For vacuum $F(R)$ gravity~(\ref{FR1}) in $D$ spacetime dimensions, the 
equation of motion is
\begin{align}
\label{JGRG13}
\frac{1}{2} \, g_{\mu\nu} F(R) - R_{\mu\nu} F'(R) - g_{\mu\nu} \Box F'(R)
+ \nabla_\mu \nabla_\nu F'(R) =0 \, .
\end{align}
If we assume constant Ricci tensor $R_{\mu\nu}$, the scalar $R$ is also 
constant and Eq.~(\ref{JGRG13}) reduces to 
\begin{align}
\label{JGRG13B}
\frac{1}{2} \, g_{\mu\nu} F(R) - R_{\mu\nu} F'(R) =0 
\end{align}
and, contracting with $g^{\mu\nu}$, one obtains the algebraic equation for 
$R$
\begin{align}
\label{JGRG13C}
\frac{D}{2} \, F(R) - R F'(R) =0\, .
\end{align}
If~(\ref{JGRG13C}) admits a positive solution $R=R_0>0$, the 
Schwarzschild-de Sitter spacetime is an exact solution while, if 
$R=R_0<0$, this is replaced by the Schwarzschild--Anti-de Sitter solution:
\begin{align}
\label{DdSdS1}
ds^2 = - f(r) dt^2 + f(r)^{-1} dr^2 + r^2 d\Omega_{(D-2)}^2 \, , \quad
f(r) = 1- \frac{2M}{r^{D-3}} - \lambda r^2 \, , \quad
R_0=D(D-1)\lambda \,.
\end{align}
where $d\Omega^2_{(D-2)}$ is the line element on the unit $(D-2)$-sphere. 
In the limit $D\to2$, the metric~(\ref{DdSdS1}) asymptotes to 
\begin{align}
\label{DdSdS5B}
ds^2 = - f(r) dt^2 + f(r)^{-1} dr^2 \, , \quad
f(r) = 1- 2Mr - \lambda r^2 \, , \quad
\lambda = \frac{R_0}{2} \, .
\end{align}
Einstein gravity in two dimensions has no black hole solutions because Eq.~(\ref{JGRG13C}) 
becomes an identity. Since $f(r)\to 1$ when $r\to 0$, the region  near the origin appears   
always flat and it seems that there cannot be singularities at $r=0$, a situation characteristic of two 
dimensions.  Moreover, since the Riemann tensor in two dimensions is constructed   
solely out of the metric and the scalar curvature, 
$R_{\mu\nu\rho\sigma}=\frac{1}{2}\left( g_{\mu\rho} g_{\nu\sigma} - g_{\mu\sigma} g_{\nu\rho} \right) R$, 
the Kretschmann scalar  looks always finite,  
$R_{\mu\nu\rho\sigma} R^{\mu\nu\rho\sigma} = R^2 = 4 \lambda^2$. However, this is not true. 
In fact, for each value of the coordinate $r$ which expresses the distance from the origin, 
there correspond two points in the  spatial dimension, that is, the region including 
one of the 
two horizons and a second region including another horizon. 
To avoid this degeneracy, we define a new coordinate $x$ by $x \equiv r\cos\theta$ 
(where $\theta=0, \pi $). Using this new coordinate, $f$ is given by 
\begin{align}
\label{fx}
f(x) = \left\{ \begin{array}{cc} 
1 - 2Mx - \lambda x^2 & \ \mbox{when}\ x \geq 0 \,,\\
1 + 2Mx - \lambda x^2 & \ \mbox{when}\ x < 0 \,.
\end{array} \right. 
\end{align}
Then a singularity appears, corresponding to a conical singularity in 
higher dimension. Due to this singularity, the scalar curvature behaves as 
\begin{align}
\label{fx2}
R(x) = 8M \delta(x) + 2 \lambda_0 \,,
\end{align}
where $\delta(x)$ is the Dirac delta function. One concludes that there is, after all, a 
curvature singularity at the origin $r=0$. 

The horizons of the geometry~(\ref{DdSdS5B}), if they exist, are located 
by the roots of $f(r)=0$ 
\begin{align}
\label{DdSdS6}
r_\pm \equiv \frac{- M \pm \sqrt{M^2 + \lambda}}{\lambda} \,.
\end{align}
In order for two distinct real and positive solutions to exist, one 
should require
\begin{align}
\label{DdSdS7}
 -M^2 < \lambda < 0 \, ,
\end{align}
which corresponds to the Schwarzschild--Anti-de Sitter (SAdS) spacetime with $R_0<0$.  
In the SAdS spacetime, $r=r_-$ is the radius of an inner black hole horizon and $r=r_+$ that 
of an outer black hole horizon.  In higher dimension, the SAdS spacetime has only one 
black hole horizon,  but in two dimensions there appear two horizons. 
The extremal limit in which the radii of these two horizons coincide corresponds to 
$M^2 = - \lambda$. When $\lambda>0$ (giving the Schwarzschild-de Sitter (SdS) spacetime with 
$R_0>0$), there is only one black hole horizon at radius $r_\mathrm{H}=r_+$.

We rewrite $f(r)$ as
\begin{align}
\label{DdSdS8}
f(r)=\frac{\left( r - r_+ \right) \left( r - r_- \right)}{r_+ \, r_-} 
\end{align}
then, when $r=r_\pm \pm \delta r$ with $\delta r/r_{\pm} \ll 1$, it is 
\begin{align}
\label{DdSdS9}
f(r)=\frac{\left( r_+ - r_- \right)\delta r}{r_+ r_-} \, .
\end{align}
Wick-rotating the time coordinate $t\to i\tau$, the metric~(\ref{DdSdS5B}) 
assumes the form 
\begin{align}
\label{DdSdS10}
ds^2=\frac{\left( r_+ - r_- \right)\delta r}{r_+ r_-} \, d\tau^2
+ \frac{r_+ r_-}{\left( r_+ - r_- \right)\delta r} \, dr^2 \, .
\end{align}
Introducing the new radial coordinate $\rho$ defined by
\begin{align}
\label{DdSdS11}
d\rho = \pm dr \sqrt{ \frac{r_+ r_-}{\left( r_+ - r_- \right)\delta r}} 
\,, \quad
 \rho = \pm 2 \sqrt{ \frac{ r_+ r_- \delta r}{r_+ - 
r_-}}\,, \quad \mbox{or} \quad \delta r= \frac{r_+ - r_-}{4 r_+ r_-}\rho^2 \, ,
\end{align}
the two-dimensional line element~(\ref{DdSdS10}) reads
\begin{align}
\label{DdSdS12}
ds^2=\frac{\left( r_+ - r_- \right)^2}{ 4 r_+^2 r_-^2} \, \rho^2d\tau^2
+ d\rho^2 \, .
\end{align}
To avoid a conical singularity in the Wick-rotated Euclidean space one 
imposes periodicity,
\begin{align}
\label{DdSdS13}
\frac{r_+ - r_-}{ 2r_+ r_-} \, \tau
\sim \frac{r_+ - r_- }{ 2r_+ r_-} \, \tau + 2\pi \, ,
\end{align}
which identifies the black hole temperature 
\begin{align}
\label{DdSdS14}
T=\frac{r_+ - r_-}{ 4\pi r_+ r_-}
= \frac{\sqrt{M^2 + \lambda}}{2\pi} 
=\pm \frac{1}{4\pi} \left( \lambda r_{\pm} +\frac{1}{r_{\pm}} \right) 
\end{align}
associated with Hawking radiation. 
Clearly, $T$ vanishes in the
extremal limit $r_+=r_-$, or $M^2=-\lambda$. 

It is difficult to define thermodynamical quantities such as entropy
in two dimensions because the area of the black hole horizon vanishes.
The area of the $\left( D-2\right)$-dimensional sphere of unit radius is 
\begin{align}
\label{DdSdS15}
A_{D-2} = \frac{ 2\pi^{\frac{D-1}{2}}}{\Gamma\left( \frac{D-1}{2} \right)} \,,
\end{align}
where $\Gamma$ is the gamma function.
When $D=2$, Eq.~(\ref{DdSdS15}) gives $A_0=2$, 
corresponding to the two points on the straight line whose center is the origin. 

If we assume $dM=dQ$, where $Q$ denotes heat, the thermodynamic definition 
of entropy $d\mathcal{S} = dQ/T$ gives
\begin{align}
\label{Ent1}
\mathcal{S}= \pi \int \frac{dM}{\sqrt{M^2 + \lambda}} = \left\{ 
\begin{array}{cc}
\mathcal{S}_0 + \pi \cosh^{-1} \left( \frac{\sqrt{-\lambda}}{M}\right) \, 
\quad & \mbox{when}\ \lambda<0 \,, \\
&\\
\mathcal{S}_0 + \pi \sinh^{-1} \left( \frac{\sqrt{\lambda}}{M} \right) \, 
\quad & \mbox{when}\ \lambda>0 \,. 
\end{array} \right. 
\end{align} 
We may identify $\mathcal{S}_0$ with the Bekenstein-Hawking entropy given 
by the ``area'' $A_{D-2=0}=2$ in (\ref{DdSdS15}), which does not 
depend 
on the black hole mass $M$ or horizon radius, but there is a 
non-trivial correction depending on $M$. For small $M$, the correction behaves as the logarithmic function 
$\cosh^{-1} \left( \frac{\sqrt{-\lambda}}{M} \right) \sim \sinh^{-1} \left( \frac{\sqrt{-\lambda}}{M} \right) \sim \ln \left( \frac{2\sqrt{\left|\lambda\right|}}{M} \right)$. 
The Bekenstein-Hawking entropy could be the leading order term in the WKB approximation, while the next-to-leading term is logarithmic. 
This suggests that the non-trivial correction in (\ref{Ent1}) corresponds to the leading quantum 
correction to the WKB approximation.

Let us discuss the relationship between the black hole geometry~(\ref{DdSdS5B}) 
in $F(R)$ gravity and the JT gravity spacetime. 
Because $R=V'(\phi)$ and  $R$ is constant for the black hole solution~(\ref{DdSdS5B}) in 
the action~(\ref{FR2}), 
$\phi$ should be a constant $\phi_0$ satisfying the equation $V'\left( \phi_0 \right) = R_0 = 2\lambda$. 
But $\e^\sigma V(\phi) = \phi \Lambda$ and the line element in the JT gravity is 
\begin{align}
\label{JTmetric}
ds^2_\mathrm{JT} = \e^{-\sigma} ds^2 = \frac{V\left( \phi_0 \right)}{\phi_0 \Lambda} 
\left[ - \left( 1- 2Mr - \lambda r^2 \right) dt^2 + \frac{dr^2}{1- 2Mr - \lambda r^2} \right] \, .
\end{align}
By rescaling radius and time as $r\to r \sqrt{\frac{V\left( \phi_0 \right)}{\phi_0 \Lambda}}$, 
$t\to t \sqrt{\frac{V\left( \phi_0 \right)}{\phi_0 \Lambda}}$, we write the line element  
(\ref{JTmetric}) as  
\begin{align}
\label{JTmetric2}
ds^2_\mathrm{JT} = - \left( 1- 2\tilde M r - \tilde \lambda r^2 \right) 
dt^2 + \frac{dr^2}{1- 2\tilde Mr - \tilde \lambda r^2} \,, 
\end{align}
where
\begin{align}
\label{tildeMlambda}
\tilde{M} = M \sqrt{\frac{V\left( \phi_0 \right)}{\phi_0 \Lambda}}\,, \quad 
\tilde{\lambda} = \frac{\lambda V\left( \phi_0 \right)}{\phi_0 \Lambda} \,.
\end{align}
The black hole mass and the effective cosmological constant are changed by constant factors 
in JT gravity. 

\subsection{Entropy in the WKB approximation}
\label{subsec2-3}

With the leading term of the entropy $\mathcal{S}$ in the WKB 
approximation constant, we estimate the leading correction in the WKB 
approximation. For the spacetime~(\ref{DdSdS5B}) with $\lambda<0$, where there are two 
horizons 
$r_\mathrm{H} \equiv r_\pm$ and 
the spacetime is asymptotically Anti-de Sitter,  the action~(\ref{FR1}) with $D=2$ becomes, 
after the Wick rotation, 
\begin{align}
\label{SAdSD3}
S_{F(R)} = \frac{1}{2\kappa^2} \int d^2 x \sqrt{-g} F(R) 
= \frac{F(R_0)}{\kappa^2} \int_0^{ 1/ T } dt \int_{r_\mathrm{H}}^L dr 
= \frac{F(R_0)\left( L - r_\mathrm{H} \right)}{\kappa^2 T}  \,,
\end{align}
where we have chosen the outer horizon $r_\mathrm{H}=r_{+}$ and we have introduced a cutoff 
$L$ to regulate the divergence of the action~(\ref{SAdSD3}). 
We consider the difference between the action of the Schwarzschild--Anti-de Sitter spacetime and that of the pure Anti-de Sitter spacetime with $M=0$. 
We determine the period $1/\tilde T$ of time in the Euclidean Anti-de Sitter space so that the physical length (between $t=0$ and $t=1/{\tilde T}$) equals the physical 
length between $t=0$ and $t= 1/ T$ in the Schwarzschild--Anti-de Sitter spacetime, 
\begin{align}
\label{SAdSB1}
\frac{ \left( 1 - 2M L - \lambda L^2 \right)^{1/2} }{T} = \frac{ \left( 1 -\lambda L^2 \right)^{1/2} }{ {\tilde T}} \, ,
\end{align}
or
\begin{align}
\label{SAdSB2}
\frac{1}{{\tilde T}} = \frac{1}{T} \sqrt{ 1 - \frac{2ML}{1 - \lambda L^2}} \sim \frac{1}{T} \left( 1 + \frac{M}{\lambda L} \right) \, .
\end{align}
Then, the action for Anti-de Sitter spacetime reads 
\begin{align}
S_\mathrm{AdS} = \frac{F(R_0)}{\kappa^2} \int_0^{1/\tilde{T} } dt 
\int_0^L dr = \frac{ F(R_0) L }{\kappa^2\tilde T} 
= \frac{ F(R_0) L }{ 2\kappa^2 T } \left( 1 + \frac{M}{\lambda L} \right) \, .
\end{align}
In the limit $L\to \infty$, the action for the black hole in the Anti-de Sitter space reduces to 
\begin{align}
\label{BH1}
S = S_{F(R)} - S_\mathrm{AdS} = - \frac{F(R_0)}{2\kappa^2 T} \left( r_\mathrm{H} + \frac{M}{\lambda} \right) 
= - \frac{F(R_0)}{2\kappa^2 T} \left( r_\mathrm{H} + \frac{1 - \lambda {r_\mathrm{H}}^2}{2 \lambda r_\mathrm{H}} \right) 
= - \frac{F(R_0)}{2\kappa^2 T} \left( \frac{r_\mathrm{H}}{2} + \frac{1}{2 \lambda r_\mathrm{H}} \right) \, ,
\end{align}
while the free energy is 
\begin{align}
\label{SAdS8}
F= - TS = \frac{F(R_0)}{4\kappa^2} \left( r_\mathrm{H} + \frac{1}{\lambda r_\mathrm{H}} \right) \, .
\end{align}
In (\ref{BH1}) we have used the definition of $r_\mathrm{H}$, 
$f\left( r_\mathrm{H} \right) = 1- 2M r_\mathrm{H} - \lambda {r_\mathrm{H}}^2 =0 $ given by Eq.~(\ref{DdSdS5B}). 
The thermodynamical relations 
\begin{align}
\label{SAdS9}
E = F - T \, \frac{d F}{d T}\, , \quad \mathcal{S} = \frac{E - F}{T} = - \frac{d F}{d T} \, ,
\end{align}
for the internal energy $E$ and the entropy $\mathcal{S}$ then provide the entropy 
\begin{align}
\label{SAdS10}
\mathcal{S} =&\, - \frac{\frac{dF}{d r_\mathrm{H}}}{\frac{d T}{d r_\mathrm{H}}}
= \frac{\frac{F(R_0)}{4\kappa^2} \left( 1 - \frac{1}{\lambda {r_\mathrm{H}}^2} \right)}
{\frac{\lambda}{4\pi}\left( 1 - \frac{1}{\lambda {r_\mathrm{H}}^2} \right)} 
= \frac{2\pi F(R_0)}{2 \lambda \kappa^2} = \frac{2\pi F'(R_0)}{\kappa^2} \, ,
\end{align}
where we have used Eq.~(\ref{DdSdS14}), Eq.~(\ref{JGRG13C}) with $D=2$, {\em i.e.}, 
$F(R) = R F'(R)$, and the last equation in (\ref{DdSdS5B}), $\lambda = R_0 / 2$. 
In general $F(R)$ gravity, $\frac{\kappa^2}{F'(R_0)}$ is regarded as the effective 
gravitational coupling. 
Equation~(\ref{SAdS10}) tells us that the entropy is constant, 
which is consistent with the large $N$ limit of the SYK model~\cite{Sarosi:2017ykf}. 
The relations between the JT gravity and the SYK model have been studied in depth \cite{Jensen:2016pah, Maldacena:2016upp}. 
The corrections to the constant entropy in (\ref{Ent1}) could be interpreted as the next-to-leading 
correction on the gravity side, or as $1/N$ corrections in the SYK model.

Several kinds of entropies have been proposed in the literature. 
The Tsallis entropy~\cite{tsallis} is 
\begin{align}
\label{TS1}
\mathcal{S}_\mathrm{T} = \mathcal{S}_0 \left( \frac{\mathcal{S}}{ \mathcal{S}_0} 
\right)^\delta \,,
\end{align}
where $ \mathcal{S}_0$ and $\delta$ are constants, while the R{\'e}nyi entropy~\cite{renyi} 
\begin{align}
\label{RS1}
\mathcal{S}_\mathrm{R}=\frac{1}{\alpha} \ln \left( 1 + \alpha \mathcal{S} \right) \, ,
\end{align}
contains the single parameter $\alpha$. 
The R\'enyi entropy was originally introduced as an index specifying the amount of information. 
In \cite{Nojiri:2022aof}, we have considered the three-parameter entropy, 
\begin{align}
\label{general6} 
\mathcal{S}_\mathrm{G} \left( \alpha, \beta, \gamma \right) 
= \frac{1}{\gamma} \left[ \left( 1  + \frac{\alpha}{\beta} \, \mathcal{S} \right)^\beta - 1 
\right] \,. 
\end{align}
The generalized entropies above are functions of the Bekenstein-Hawking entropy $\mathcal{S}$, which is a constant, and therefore 
they are also constant and do not contribute to the dynamics of the system. 

\section{Dynamical dilaton gravity in two dimensions}\label{Sec3}

In the case of $D=2$~ $F(R)$ gravity and of JT gravity, the thermal entropies of the black 
hole, including the Bekenstein-Hawking entropy and the generalized entropies, are constant. 
These constants do not depend on the black hole radius nor the Hawking temperature. 
In this section, we consider dilaton gravity with a kinetic term  (which we refer to as 
``dynamical dilaton gravity'') and we show that 
the thermal entropy can be a general function of the black hole radius and the Hawking temperature 
(see \cite{Nojiri:2000ja} for a review of dilaton gravity).

\subsection{Construction of a black hole spacetime}\label{subsec3-1}

Following Ref.~\cite{Nojiri:2022sfd}, we first recall the basics of dilaton gravity with a 
kinetic term and then show how to construct an 
arbitrary static black hole geometry in this framework.

The action of dilaton gravity in two dimensions is 
\begin{align}
\label{dgbh1}
S= \int d^2 x \sqrt{-g} \left[ \phi R - \frac{\omega(\phi)}{2} \, 
\partial_\mu \phi \partial^\mu \phi - V(\phi) \right] \,.
\end{align}
By varying the action (\ref{dgbh1}) with respect to $\phi$, one obtains 
\begin{align}
\label{dgbh2}
R +  \frac{\omega'(\phi)}{2} \, \partial_\mu \phi \partial^\mu \phi + \omega(\phi) \, \Box 
\phi - V'(\phi) =0 \,,
\end{align}
while the variation with respect to the inverse metric $g^{\mu\nu}$ yields 
\begin{align}
\label{dgbh3}
\frac{1}{2} \, g_{\mu\nu} \left[ \phi R - \frac{\omega(\phi)}{2} \partial_\rho \phi \partial^\rho \phi - V(\phi) \right] 
 - \phi R_{\mu\nu} + \nabla_\mu \nabla_\nu \phi - g_{\mu\nu} \Box \phi + \frac{\omega(\phi)}{2} \, \partial_\mu \phi \partial_\nu \phi =0 \, .
\end{align}
Let the coordinates in the two-dimensional spacetime be $\left( t, r \right)$. 
There are two gauge degrees of freedom corresponding to 
general coordinate transformations and we can choose 
\begin{align}
\label{dgbh4}
g_{tr}=g_{rt}=0\, , \quad\quad g_{tt} \, g_{rr} = -1 \, , 
\end{align}
yielding the line element 
\begin{align}
\label{dgbh5}
ds^2 = - \e^{2\nu} dt^2 + \e^{-2\nu}dr^2 \, .
\end{align}
Searching for static geometries, we assume that $\nu$ and $\phi$ depend only on the radial coordinate $r$. 
Then the connection is  
\begin{align}
\label{dgbh6}
\Gamma^t_{tt} = \Gamma^t_{rr} = \Gamma^r_{tr} = \Gamma^r_{rt} = 0\, , 
\quad \Gamma^t_{tr} = \Gamma^t_{rt} = \e^{-4\nu} \Gamma^r_{tt} 
= - \Gamma^r_{rr} = \nu'
\end{align}
and the Ricci tensor and Ricci scalar 
\begin{equation}
R_{\mu\nu} \equiv - \Gamma^\rho_{\mu\rho,\nu} + \Gamma^\rho_{\mu\nu,\rho} 
 - \Gamma^\sigma_{\mu\rho} \Gamma^\rho_{\nu\sigma} + \Gamma^\sigma_{\mu\nu} \Gamma^\rho_{\sigma\rho} \,, \quad 
R \equiv g^{\mu\nu} R_{\mu\nu} \,,
\end{equation}
are computed as 
\begin{align}
\label{dgbh7}
R_{tt} = &\, \e^{4\nu} \left( \nu'' + 2 {\nu'}^2 \right) \, , \quad 
R_{tr} = R_{rt} = 0 \, , \quad R_{rr} = - \nu'' - 2 {\nu'}^2 \, , 
\nonumber \\
R= &\, - 2 \e^{2\nu} \left( \nu'' + 2 {\nu'}^2 \right) \,.
\end{align}
These expressions provide the explicit form of Eqs.~(\ref{dgbh2}) and (\ref{dgbh3}), 
\begin{align}
\label{dgbh8}
0=&\, - 2 \e^{2\nu} \left( \nu'' + 2 {\nu'}^2 \right) + \frac{1}{2} \e^{2\nu} {\phi'}^2 
+ \omega(\phi) \e^{2\nu} \left(\phi'' + 2 \nu' \phi' \right) - V'(\phi) \, , \\
\label{dgbh9}
0=&\, - \frac{\e^{2\nu}}{2} \left[ - \frac{ \omega(\phi)}{2} \e^{2\nu} {\phi'}^2 - V(\phi) \right] 
+ \e^{4\nu} \left(\phi'' - 2 \nu' \phi' \right) \, , \\
\label{dgbh10}
0=&\, \frac{ \e^{-2\nu} }{2} \left[ \frac{ \omega(\phi)}{2} \e^{2\nu}{\phi'}^2 - V(\phi) \right] - \nu' \phi' \, .
\end{align}
If Eqs.~(\ref{dgbh9}) and (\ref{dgbh10}) are satisfied, then also Eq.~(\ref{dgbh8}) is automatically satisfied. 
Equations~(\ref{dgbh9}) and (\ref{dgbh10}) can be solved with respect to $V(\phi)$ and $\omega(\phi)$, obtaining 
\begin{align}
\label{dgbh11}
V(\phi) = \e^{2\nu} \left( - \phi'' + \nu' \phi' \right) \, , \quad 
\omega(\phi) = - 2 \phi'' + 6 \nu' \phi' \, .
\end{align}
Let us assume $\nu$ and $\phi$ to be arbitrary functions of $r$, $\nu=\nu(r)$ 
and $\phi=\phi(r)$; the equation $\phi=\phi(r)$ can be inverted (at least locally) to obtain 
$r=r(\phi)$. If $V(\phi)$ and $\omega(\phi)$ are given by 
\begin{align}
\label{dgbh12}
V(\phi) = \e^{2\nu}\left( r\left(\phi\right) \right) 
\left[ - \phi''\left( r\left(\phi\right)\right) + \nu'\left( r\left(\phi\right)\right) 
\phi'\left( r\left(\phi\right)\right) \right] \, , \quad 
\omega(\phi) = - 2 \phi''\left( r\left(\phi\right)\right) + 6 \nu'\left( r\left(\phi\right)\right) \phi'\left( r\left(\phi\right)\right) \, ,
\end{align}
an arbitrary configuration $\nu=\nu(r)$, $\phi=\phi(r)$ can be 
realized in this model. 

Using Eq.~(\ref{dgbh9}), the action~(\ref{dgbh1}) is rewritten as 
\begin{align}
\label{dgbh13}
S =&\, \int d^2x \left[ - 2 \phi \, \e^{2\nu} \left( \nu'' + 2 {\nu'}^2 
\right) - \frac{ \omega(\phi)}{2} \, \e^{2\nu}{\phi'}^2 - V(\phi) \right] 
= \int d^2x \left\{ -2 \left(\phi' \e^{2\nu}\right)' + \left[ \phi \left(\e^{2\nu}\right)'\right]' \right\} \, ,
\end{align}
which will be used below to evaluate the thermodynamical quantities. 
We should note that the situation is different from the entropy in four-dimensional spacetime, where the consistency of entropy with the Hawking temperature 
selects the Bekenstein-Hawking entropy as the unique consistent entropy for black holes (see \cite{Nojiri:2022aof, Nojiri:2022sfd}). 

\subsection{Example}\label{subsec3-2}

For illustration, consider the Schwarzschild--(Anti-)de 
Sitter spacetime 
$\e^{2\nu}=f(r) = 1- 2Mr - \lambda r^2$ in (\ref{DdSdS5B}) and assume 
\begin{align}
\label{dgbh14}
\phi=C\left( 1 - \frac{{r_0}^2}{{r_0}^2 + r^2} \right) \,,
\end{align}
where $r_0$ is a positive constant 
(there is no particular physical motivation for this choice that we make to provide an explicit example). 
The substitution of Eqs.~(\ref{DdSdS5B}) and (\ref{dgbh14}) into Eq.~(\ref{dgbh12}) yields 
\begin{align}
\label{dgbh15}
V(\phi) =&\, \left\{ 1 - 2Mr_0\sqrt{ \frac{1}{1 - \frac{\phi}{C}} - 1}
 - \lambda {r_0}^2 \left( \frac{1}{1 - \frac{\phi}{C}} - 1\right)  \right\} \left\{ \frac{6C}{{r_0}^2} \left( 1 - \frac{\phi}{C} \right)^2 
 + \frac{8C}{{r_0}^2} \left( 1 - \frac{\phi}{C} \right)^3 \right\} \nonumber \\
&\, - \left\{ M + \lambda r_0\sqrt{ \frac{1}{1 - \frac{\phi}{C}} - 1} \right\}
\frac{2C}{r_0}\left( 1 - \frac{\phi}{C} \right)^2 \sqrt{ \frac{1}{1 - \phi/C } - 1} \,, \nonumber \\
\omega(\phi) =&\, 2 \left\{ \frac{6C}{{r_0}^2} \left( 1 - \frac{\phi}{C} \right)^2 
+ \frac{8C}{{r_0}^2} \left( 1 - \frac{\phi}{C} \right)^3 \right\} 
+ \frac{\frac{6C}{r_0} \left(M + \lambda r_0\sqrt{ \frac{1}{1 - \frac{\phi}{C}} - 1}\right) 
\left( 1 - \frac{\phi}{C} \right)^2 \sqrt{ \frac{1}{1 - \frac{\phi}{C}} - 1}}
{1 - 2Mr_0\sqrt{ \frac{1}{1 - \frac{\phi}{C}} - 1} - \lambda {r_0}^2 \left( \frac{1}{1 - \frac{\phi}{C}} - 1\right)} \, .
\end{align}
The expressions in~(\ref{dgbh15}) include the mass parameter $M$, which 
looks unphysical because the mass usually appears as an integration constant. 
To avoid this problem we add to the action~(\ref{dgbh1}) the term\footnote{A similar prescription has been used in Ref.~\cite{Nojiri:2017kex}.} 
\begin{align}
\label{dgbh16}
S_{\xi\eta} = \int d^2 x \sqrt{-g} \, \xi^\mu \partial_\mu \eta 
\end{align}
with a vector field $\xi^\mu$ and a scalar field $\eta$, and we replace 
$M$ in Eq.~(\ref{dgbh15}) with $\eta$, 
\begin{align}
\label{dgbh17}
V(\phi) =&\, \left\{ 1 - 2\eta r_0\sqrt{ \frac{1}{1 - \frac{\phi}{C}} - 1} - \lambda {r_0}^2 \left( \frac{1}{1 - \frac{\phi}{C}} - 1\right) \right\}
\left\{ \frac{6C}{{r_0}^2} \left( 1 - \frac{\phi}{C} \right)^2 + \frac{8C}{{r_0}^2} \left( 1 - \frac{\phi}{C} \right)^3 \right\} \nonumber \\
&\, - \left\{ \eta + \lambda r_0\sqrt{ \frac{1}{1 - \frac{\phi}{C}} - 1} \right\}
\frac{2C}{r_0}\left( 1 - \frac{\phi}{C} \right)^2 \sqrt{ \frac{1}{1 - \frac{\phi}{C}} - 1} \nonumber \\
\omega(\phi) =&\, 2 \left\{ \frac{6C}{{r_0}^2} \left( 1 - \frac{\phi}{C} \right)^2 
+ \frac{8C}{{r_0}^2} \left( 1 - \frac{\phi}{C} \right)^3 \right\} 
+ \frac{\frac{6C}{r_0} \left(\eta + \lambda r_0\sqrt{ \frac{1}{1 - \frac{\phi}{C}} - 1}\right) 
\left( 1 - \frac{\phi}{C} \right)^2 \sqrt{ \frac{1}{1 - \frac{\phi}{C}} - 1}}
{1 - 2\eta r_0\sqrt{ \frac{1}{1 - \frac{\phi}{C}} - 1}
 - \lambda {r_0}^2 \left( \frac{1}{1 - \frac{\phi}{C}} - 1\right)} \, .
\end{align} 
The variation of the new action~(\ref{dgbh16}) (which is the only one 
that depends on $\xi^\mu$) with respect to $\xi^\mu$ yields 
\begin{align}
\label{dgbh18}
0=\partial_\mu \eta\, ,
\end{align}
then $\eta$ is constant and we may identify it with the mass, $\eta=M$. 
When we evaluate the thermodynamical quantities, the action~(\ref{dgbh16}) 
does not contribute because $S_{\xi\eta}$ vanishes on the solutions of the 
field equation~(\ref{dgbh18}). 

\subsection{Thermodynamical quantities}\label{subsec3-3}

We now calculate the thermodynamical quantities by evaluating the action~(\ref{dgbh13}) in 
the case $\lambda<0$, 
\begin{align}
\label{dgbh19}
S =&\,  \int_0^{1/T} dt \int_{r_\mathrm{H}}^L dr \left\{ -2 \left(\phi' \e^{2\nu}\right)' 
+ \left[  \phi \left(\e^{2\nu}\right)'\right]' \right\}
= \frac{1}{T}  \left[ -2 \phi' \e^{2\nu} + \phi \left(\e^{2\nu}\right)' 
\right]_{r_\mathrm{H}}^L \nonumber \\
\sim &\, \frac{1}{T} \left[ - 2 C \lambda L + C\left( 1 - \frac{{r_0}^2}{{r_0}^2 
+ {r_\mathrm{H}}^2} \right)\left( \frac{1}{r_\mathrm{H}} + \lambda r_\mathrm{H} \right) 
\right] \, .
\end{align}
Here we have used the definition of the horizon radius $ 1 - 2M r_\mathrm{H} - \lambda {r_\mathrm{H}}^2 =0 $, $L$ is the cutoff, 
$T$ is the Hawking temperature~(\ref{DdSdS14}), and $r_\mathrm{H}$ is the outer horizon radius in (\ref{DdSdS6}) $\left(r_\mathrm{H}=r_+\right)$. 
We consider again the difference between the action of the Schwarzschild--Anti-de Sitter spacetime in (\ref{dgbh19}) and that of the 
pure Anti-de Sitter spacetime with $M=0$. 
The Euclidean time period $ 1 / \tilde T $ in Anti-de Sitter spacetime is given by (\ref{SAdSB2}), then the action $S_\mathrm{AdS}$ 
for Anti-de Sitter spacetime is evaluated as
\begin{align}
\label{dgbh20}
S_\mathrm{AdS} =&\, \int_0^{ 1/\tilde T } dt \int_0^L dr \left[ -2 \left(\phi' \e^{2\nu}\right)' + \left( \phi \left(\e^{2\nu}\right)'\right)' \right] 
\sim - \frac{2 C \lambda L}{T} \left( 1 + \frac{M}{\lambda L} \right) \,. 
\end{align}
In the limit $L\to \infty$, we obtain the free energy 
\begin{align}
\label{dgbh21}
F= -T \left( S - S_\mathrm{AdS} \right) 
= - C \left\{ \frac{2}{r_\mathrm{H}} - \frac{{r_0}^2}{{r_0}^2 + {r_\mathrm{H}}^2} \left( \frac{1}{r_\mathrm{H}} 
+ \lambda r_\mathrm{H} \right) \right\} \, .
\end{align}
By using the thermodynamical relations (\ref{SAdS9}), we find the entropy 
\begin{align}
\label{dgbh22}
\mathcal{S} = - \frac{\frac{dF}{d r_\mathrm{H}}}{\frac{d T}{d r_\mathrm{H}}} 
= \frac{ C\left\{ - \frac{2}{{r_\mathrm{H}}^2} + \frac{2\lambda {r_0}^2 {r_\mathrm{H}}^2}{\left({r_0}^2 + {r_\mathrm{H}}^2\right)^2} 
\left( 1 + \frac{1}{{\lambda r_\mathrm{H}}^2} \right) - \frac{\lambda {r_0}^2}{{r_0}^2 + {r_\mathrm{H}}^2} 
\left( 1 - \frac{1}{{\lambda r_\mathrm{H}}^2} \right) \right\}}{\frac{\lambda}{2}\left( 1 - \frac{1}{\lambda {r_\mathrm{H}}^2} \right)} \, , 
\end{align}
which is a non-trivial function of the horizon radius $r_\mathrm{H}$.

To recap, we have evaluated the entropy for a specific model with 
action given by the sum of~(\ref{dgbh1}) with (\ref{dgbh17}) and the action (\ref{dgbh16}). 
We can, however, calculate the entropy for the general model  with arbitrary $V(\phi)$ and 
$\omega(\phi)$ by using the 
expression (\ref{dgbh13}), where the Lagrangian density is a total derivative,
\begin{align}
\label{dgbh13}
S = - \frac{F}{T} = \frac{1}{T} \left[ -2 \phi' \e^{2\nu} + \phi \left(\e^{2\nu}\right)' 
\right]_{r_\mathrm{H}}^L   - \frac{1}{\tilde T} \left[ -2 \phi' \e^{2\nu} + \phi 
\left(\e^{2\nu}\right)' \right]_{0}^L \, .
\end{align}
Therefore the entropy becomes a general function of the black hole mass $M$, the horizon radius $r_\mathrm{H}$, or the Hawking temperature $T$, 
which depends on the form of $V(\phi)$ and $\omega(\phi)$. 
If we assume the Schwarzschild--Anti-de Sitter spacetime in 
(\ref{DdSdS5B}), by using (\ref{SAdSB2}), the free energy $F$ in (\ref{dgbh13}) is expressed as 
\begin{align}
\label{dgbh14BB}
F =&\, \left[ 2 \phi' (L) \left( 1- 2ML - \lambda L^2 \right) + 2 \phi (L) \left( M + \lambda L \right) \right]
 - \left[ 2 \phi'\left(r_\mathrm{H} \right) \left( 1- 2Mr_\mathrm{H} - \lambda {r_\mathrm{H}}^2 \right) 
+ 2 \phi \left(r_\mathrm{H}\right) \left( M + \lambda r_\mathrm{H} \right) \right] \nonumber \\
&\, - \left( 1 + \frac{M}{\lambda L} \right) \left\{ \left[ 2 \phi' (L) \left( 1- 2ML - \lambda L^2 \right) + 2 \phi (L) \left( M + \lambda L 
\right) \right] - \left[ 2 \phi' (0) + 2 M \phi (0) \right] \right\} \, .
\end{align}
We may assume that $\phi(0)=\phi'(0)=0$ and 
$\phi(r)\sim \phi_0 + \frac{\phi_1}{r} +\mathcal{O}\left( r^{-2} \right)$ as $r\to \infty$, where  $\phi_0$ and $\phi_1$ are constants. 
Then the limit $L\to\infty$ yields
\begin{align}
\label{dgbh15BB}
F \to - \phi \left(r_\mathrm{H}\right) \left( \frac{1}{r_\mathrm{H}} + \lambda r_\mathrm{H} \right) -2M \phi_0\,,
\end{align}
where we have used again the definition of horizon radius 
$ 1 - 2M r_\mathrm{H} - \lambda {r_\mathrm{H}}^2 =0 $. 
The entropy is 
\begin{align}
\label{dgbh16BB}
\mathcal{S} = \frac{2}{\lambda - \frac{1}{{r_\mathrm{H}}^2}} 
\left[ \phi \left(r_\mathrm{H}\right) \left( \frac{1}{r_\mathrm{H}} + \lambda r_\mathrm{H} \right) \right]' \, .
\end{align}
According to this equation, if we choose $\phi(r)$ by using a function $f(r)$ such that, 
when $r\sim r_\mathrm{H}$, 
\begin{align}
\label{dgbh17}
\phi(r)=\frac{1}{2 \left(\frac{1}{r} + \lambda r \right)} \int_{r_0}^r dr \left( \lambda - \frac{1}{r^2} \right) f(r)\, ,
\end{align}
then 
\begin{align}
\label{dgbh18BB}
\mathcal{S}=f\left(r_\mathrm{H} \right) \, .
\end{align}
In (\ref{dgbh17}), $r_0$ is a constant, which gives an integration constant. 
Because the function $f(r)$ is (almost) arbitrary, we can obtain an entropy which is an arbitrary function of the horizon radius 
$r_\mathrm{H}$, which in turn can be rewritten in terms of the black hole mass $M$ or the Hawking temperature $T$. 
For example, as an analog of the three-parameter generalized entropy (\ref{general6}), we consider the four-parameter entropy
\begin{align}
\label{dgbh19}
\mathcal{S} \left( \tilde\alpha, \tilde\beta, \tilde\gamma, \tilde\delta \right) 
= \frac{1}{\tilde\gamma} \left[ \left( 1 + \frac{\tilde\alpha}{\tilde\beta} \, {r_\mathrm{H}}^{\tilde\delta} 
\right)^\beta - 1 \right]
\end{align}
by replacing $\left( \alpha,\beta,\gamma, \mathcal{S} \right)$ in  (\ref{general6}) with 
$\left( \tilde\alpha, \tilde\beta, \tilde\gamma, {r_\mathrm{H}}^{\tilde\delta} \right)$. 
Because we are considering the Schwarzschild--Anti-de Sitter spacetime~(\ref{DdSdS5B}) with 
$\lambda>0$,  by using (\ref{SAdSB2}) we solve (\ref{DdSdS14}) with respect to 
$r_\mathrm{H}=r_+$, obtaining 
\begin{align}
\label{dgbh20}
r_\mathrm{H}=\frac{1}{\lambda} \left( 2\pi T \pm \sqrt{ 4\pi^2 T^2 - \lambda} \, \right)\, .
\end{align}
Physically, a smaller black hole should have a higher temperature, hence 
we choose the positive sign in~(\ref{dgbh20}), which leads to
\begin{align}
\label{dgbh21}
r_\mathrm{H}=\frac{1}{\lambda} \left( 2\pi T + \sqrt{ 4\pi^2 T^2 - \lambda} \, \right)\, .
\end{align}
By using (\ref{dgbh21}), we rewrite Eqs.~(\ref{dgbh18}) and (\ref{dgbh19}) 
as functions of the Hawking temperature $T$ as 
\begin{align}
\mathcal{S}=&\, f\left( \frac{1}{\lambda} \left( 2\pi T + \sqrt{ 4\pi^2 T^2 - \lambda} \right) \right) \, , \\
\label{dgbh23}
\mathcal{S} \left( \tilde\alpha, \tilde\beta, \tilde\gamma, \tilde\delta \right) 
=&\, \frac{1}{\tilde\gamma} \left[ \left( 1 + \frac{\tilde\alpha}{\tilde\beta} \left\{ \frac{1}{\lambda} \left( 2\pi T 
+ \sqrt{ 4\pi^2 T^2 - \lambda} \right) \right\}^{\tilde\delta} \right)^\beta - 1 \right] \, , 
\end{align}
obtaining the entropies as definite thermodynamical quantities. 

We now consider the relation between the mass $M$ and the thermodynamical energy $E$ in (\ref{SAdS9}). 
By combining (\ref{dgbh15BB}) and (\ref{dgbh17}), we find 
\begin{align}
\label{dgbh24}
F = - \frac{1}{2} \int_{r_0}^{r_\mathrm{H}} dr \left( \lambda - \frac{1}{r^2} \right) f(r) -2M \phi_0\, .
\end{align}
By using the expression of the temperature $T$ in Eq.~(\ref{DdSdS14}) and the definition of the horizon radius 
$r_\mathrm{H}$, $M=\frac{1}{2} \left( - \lambda r_\mathrm{H} + \frac{1}{r_\mathrm{H}} \right)$,  
\begin{align}
\label{dgbh25}
\frac{dT}{dr_\mathrm{H}}=\frac{1}{4\pi} \left( \lambda - \frac{1}{{r_\mathrm{H}}^2} \right)\, , \quad 
\frac{dM}{dr_\mathrm{H}}= - \frac{1}{2} \left( \lambda + \frac{1}{{r_\mathrm{H}}^2} \right) \, ,
\end{align}
by using (\ref{SAdS9}), we find 
\begin{align}
\label{dgbh26}
E = - \frac{1}{2} \int_{r_0}^{r_\mathrm{H}} dr \left( \lambda - \frac{1}{r^2} \right) f(r)  
+ \frac{1}{2} \left( \lambda r_\mathrm{H} + \frac{1}{r_\mathrm{H}} \right) f  \left( 
r_\mathrm{H} \right)  - \frac{ 4 \lambda }{\lambda r_\mathrm{H} 
- \frac{1}{r_\mathrm{H}}} \, \phi_0 
\end{align}
which, in general, is different from the mass  $M=\frac{1}{2} \left( 
\frac{1}{r_\mathrm{H}} - 
\lambda r_\mathrm{H} \right)$. 
The discrepancy could be generated by the effective energy of the scalar field. 
Birkhoff's theorem does not hold in two dimensions, therefore there is no reason for the 
mass $M$ to coincide with the thermodynamical energy. 

Since the entropy was derived in the WKB approximation, it can be regarded as the analog of 
the Bekenstein-Hawking entropy although it is not constant. Therefore, this entropy may 
change in certain situations where quantum effects dominate. By replacing the 
Bekenstein-Hawking entropy with the expression~(\ref{dgbh18}) in the Tsallis entropy 
(\ref{TS1}), the R{\'e}nyi entropy (\ref{RS1}), the Kaniadakis entropy, in the non-extensive 
entropy of the Loop Quantum Gravity, the three-parameter generalized entropy~(\ref{general6}) 
or the six-parameter entropy, one obtains a new class of generalized entropies. However, 
Eq.~(\ref{dgbh18}) implies that, even at the classical level or in the WKB approximation, 
entropies with more general forms could appear.

\section{Discussion and conclusions}\label{Sec4}

Given that the Bekenstein-Hawking black hole entropy is proportional to its horizon area and 
that this horizon degenerates into two disconnected points (with zero areas) in two spacetime 
dimensions, the Bekenstein-Hawking entropy becomes a constant independent of the black hole 
mass $M$, horizon radius $r_\mathrm{H}$, or the Hawking temperature $T$. This fact could be 
consistent with Einstein's gravity becoming topological in two dimensions. In this work, we 
have shown (in Eq.~(\ref{SAdS10})) that the entropy is constant also for $F(R)$ gravity, 
which is non-trivial in two dimensions, by evaluating explicitly its action. Since 
two-dimensional $F(R)$ gravity is equivalent to JT gravity (as shown in 
\cite{Nojiri:2022mfi}), the black hole entropy of JT gravity is also constant. By using 
the AdS/CFT correspondence, JT gravity has been shown to be equivalent to the SYK model. 
For this model, it is known that the thermal entropy is constant in the large $N$ limit 
\cite{Sarosi:2017ykf}, which is consistent with the result reported here. 
Equation~(\ref{Ent1}), however, suggests that there could be non-trivial corrections 
depending on the horizon radius or the Hawking temperature. Such corrections might arise from 
the next-to-leading order corrections in the WKB approximation in the black hole side, or 
from $1/N$ corrections in the SYK model, an interesting view to develop in future 
research.

New kinds of entropies have been proposed recently, including the Tsallis entropy 
(\ref{TS1}), the R{\'e}nyi entropy (\ref{RS1}), the Kaniadakis entropy, the non-extensive entropy 
of Loop Quantum Gravity, or the generalized entropies (\ref{general6}) with three 
or six parameters proposed by us. These new entropies are functions of the Bekenstein-Hawking 
entropy, therefore they all reduce to physically irrelevant constants in $D=2$ spacetime 
dimensions. If $D\neq 2$, however, the Bekenstein-Hawking entropy is a function of black hole 
mass $M$, horizon radius $r_\mathrm{H}$, or Hawking temperature $T$ and the above-mentioned 
generalized entropies can be regarded as general functions of these quantities. This 
observation suggests that, if they exist, generalized entropies in $D=2$ could be general 
functions of $M$, $r_\mathrm{H}$, $T$. In fact, in $D=2$ dilaton gravity, the black hole 
entropies are, in general, not constant, and in the CGHS model, the Wald entropy is different 
from the Bekenstein-Hawking one and depends on the value of the dilaton. Motivated by this 
observation, in this work we have considered general dilaton gravity and we have shown that 
one can obtain solutions describing arbitrary static spacetimes. For the models studied, the 
entropies $\mathcal{S}$ are general functions of $M$, $r_\mathrm{H}$, $T$ as in 
(\ref{entropy}) and they could be new types of generalized entropies. We provided an explicit 
example with action given by the sum of the action (\ref{dgbh1}) with (\ref{dgbh17}) and the 
action (\ref{dgbh16}) and we have shown that the corresponding entropy does indeed depend on 
$M$, $r_\mathrm{H}$, $T$ and is not constant. Although explicit calculations have only been 
given for this example, Eq.~(\ref{dgbh13}) shows that it is possible to obtain non-trivial 
entropies that are general functions of $M$, $r_\mathrm{H}$, $T$.

\section*{Acknowledgments}

This work is supported, in part, by MINECO (Spain), project PID2019-104397GB-I00 (S.~D.~O.) 
and by the Natural Sciences \& 
Engineering Research Council of Canada, Grant No.~2016-03803 (V.~F.).

\end{document}